# Gyroid-like metamaterials: Topology optimization and Deep Learning

Asha Viswanath[1,2], Diab W Abueidda[4,5], Mohamad Modrek[1,3], Kamran A Khan[1,2*], Seid Koric[4,5], Rashid K. Abu Al-Rub[1,3]

[1]Advanced Digital & Additive Manufacturing Center, Khalifa University of Science and Technology, Abu Dhabi, PO Box 127788, United, Arab Emirates,
[2]Department of Aerospace Engineering, Khalifa University of Science and Technology, Abu Dhabi, PO Box 127788, United, Arab Emirates,
[3]Department of Mechanical Engineering, Khalifa University of Science and Technology, Abu Dhabi, PO Box 127788, United, Arab Emirates,
[4]Department of Mechanical Science and Engineering, University of Illinois at Urbana-Champaign, United States of America,
[5]National Center for Supercomputing Applications, University of Urbana-Champaign, IL 61801, USA

*Corresponding author: kamran.khan@ku.ac.ae

## ABSTRACT

Triply periodic minimal surface (TPMS) metamaterials characterized by mathematically-controlled topologies exhibit better mechanical properties compared to uniform structures. The unit cell topology of such metamaterials can be further optimized to improve a desired mechanical property for a specific application. However, such inverse design involves multiple costly 3D finite element analyses in topology optimization and hence has not been attempted. Data-driven models have recently gained popularity as surrogate models in the geometrical design of metamaterials. In this paper, we build a deep learning based surrogate model for the topology optimization of a Schoen's Gyroid TPMS unit cell to obtain the optimal 3D TPMS unit cell topology for desired properties without requiring intensive computation. Gyroid-like unit cells are designed using a novel voxel algorithm, a homogenization-based topology optimization, and a Heaviside filter to attain optimized densities of 0-1 configuration. Few optimization data are used as input-output for supervised learning of the topology optimization process from a 3D CNN model. These models could then be used to instantaneously predict the optimized unit cell geometry for any topology parameters, thus alleviating the need to run any topology optimization for future design. The high accuracy of the model was demonstrated by a low mean square error metric and a high dice coefficient metric. This accelerated design of 3D metamaterials opens the possibility of designing any computationally costly problems involving complex geometry of metamaterials with multi-objective properties or multi-scale applications.

1. **INTRODUCTION**

Metamaterials have emerged in the recent past as a 'holy grail' to material scientists as they showed abundant possibilities in their physical properties and the versatility in the fields of applications (mechanical, thermal, acoustic, optical, electromagnetics, bio-medical to name a few [1],[2],[3],[4],[5]). Their mechanical properties studied by engineers showed colossal promise as their unique architectures, which could be tailored to any desired geometry, enhanced the properties of the structure beyond the capabilities of the material [6],[7],[8]. The attractive feature was that their extreme properties could be topologically controlled.

The microstructure of the base unit of these materials, referred to as the representative unit cell (RUC), determines their mechanical and physical properties [9]. The design of the RUC of metamaterials satisfying some desired properties is called the 'inverse design' problem and has been performed through experiments and/or topology optimization (TO) [10],[11]. TO aims to obtain optimal layouts of the microstructure for a desired objective function of a metamaterial such as maximizing the bulk/shear moduli or minimizing Poisson's ratio, subject to constraints, such as volume constraint [12]. This area of research has been extensively studied in 2D [13],[14] and 3D microstructures and MATLAB codes are also available for the same [15]. An initial design of the microstructure may or may not be used. Some of the initial designs used in the literature consist of simple designs with a hole at the center or a few distributed voids, which after topology optimization, give new topologies satisfying the desired objective [14].

Triply periodic minimal surface (TPMS), a concept from differential geometry, is one of the topologies adopted for the RUC of micro-structured materials. These surfaces minimize the surface area locally for a given boundary and possess the property of the mean curvature being zero at every point on the surface [16]. They divide the unit cell domain into two or more non-intersecting domains. What makes them attractive is their fascinating topologies, when repeated periodically in 3D. TPMS can be mathematically-controlled and exhibit some unique properties, such as a large surface area to volume ratio [17]. The advances in the manufacturing industry, like the use of additive manufacturing [18], also facilitated their fabrication, which was previously a major inhibition in their usage with traditional methods. Many research works have dealt with designs of the TPMS structures based on experimental studies on its properties due to its geometry [19],[20],[21],[22],[23],[24]. This work deals with a specific TPMS of the Gyroid structure. Gyroid TPMS structure is used in various applications including orthopedic implants due to its efficient load transfer along with continuous filling of the void space [25] and catalytic converters due to efficient heat transfer through void space [26]. Gyroid-structure is also found in nature in soap films [27] and butterfly wings [28]. In this work, we propose a novel method of designing 'Gyroid-like' unit cells for a desirable mechanical property subject to boundary conditions and a volume constraint using TO. Here, we start with the Gyroid structure as our initial design, then optimize it for a specific objective function yielding a Gyroid-like structure but with optimized properties. The novel approach discussed in this work captures the surface geometry of TPMS in a voxel form, and when subjected to TO, it renders a design similar to Gyroid but which may not possess the property of mean curvature being zero at all points. In other words, an initial design of a voxelized Gyroid isosurface is subjected to TO to obtain a Gyroid-like final structure with improved material distribution satisfying the desired objective and the volume constraints.

The major challenge faced during 3D unit-cell design using the above approach is the computational time taken for TO, which exponentially increases with the number of finite elements or the mesh size (number of voxels in this study) of the unit cell. For example, a mesh of 32 elements in all three dimensions takes around 67seconds on a Workstation for a single iteration of the optimization process, which may take around 200-800 iterations to converge! To alleviate this cost of computations, we search for an alternative model to the optimization process that can use information from a few optimization runs and can consequently be used as a computationally cheap alternative for unit cell design. Recently machine learning models have emerged as surrogate models to ease the computationally intensive design and make possible the design even on a laptop. Among them, many references in literature on deep learning models ([29],[30],[31] [32],[33],[34],[35],[36],[37] ,[38],[39]) inspired the authors to use them in this context of TO for designing TPMS based metamaterials. Deep learning models based on 2D convolutional neural networks (CNN) have been used in literature for this purpose in 2D unit cell generation [40], [41],[42]. CNNs are found to be robust in image recognition tasks, and this advantage of CNN is exploited for quantitatively predicting mechanical properties of composite structures over the entire volume fraction space by using checkerboard composites as image inputs to CNN [42]. Inspired by these works, we extend these CNN-based models to predict 3D unit cell TO design. Optimizing a TPMS geometry for designing the unit cell of metamaterials using TO to attain the desired objectives has not been attempted in literature other than by the authors themselves [43]. This study improves from this previous work by authors in two ways – 1) previous work dealt only with a single topology optimization parameter of volume fraction and objective function of bulk modulus and 2) The optimal densities did not follow a 0-1 configuration and hence not learned well by the CNN algorithm thereby showing high mean square error for the CNN model. This study focuses on this gap in the previous work and proposes the potential applications of using such accelerated 3D TO for unit cell design of Gyroid TPMS.

The structure of the paper is as follows: Section 2 explains the novel methodology of generation of Gyroid-like structures for unit cell geometry and design using 3D homogenization based TO. Section 3 elaborates the surrogate deep learning model describing the data generation procedure and the architecture of the 3D CNN network used. Section 4 lists the different errors encountered in various approximations in this study and how they are accounted for. Section 5 discusses the results obtained from the proposed model. The last section summarizes the insights gained from this model and future directions in this research.

## 2. METHODOLOGY

### 2.1 Voxel-based architecture of Gyroid TPMS

The generation of the Gyroid microstructure is first discussed. Figure 1a shows the isosurface of a Gyroid structure along with its unit cell and periodic structure (Figure 1b). This surface is generated from the level set approximation equation

$$\sin\frac{2\pi x}{L_x}\cos\frac{2\pi y}{L_y}+\sin\frac{2\pi y}{L_y}\cos\frac{2\pi z}{L_z}+\sin\frac{2\pi z}{L_z}\cos\frac{2\pi x}{L_x}=c \quad (1)$$

where $c$ denotes level set value which can be a constant or a function of $x, y,$ and $z$. $L_x$, $L_y,$ and $L_z$ are unit cell lengths in the three directions.

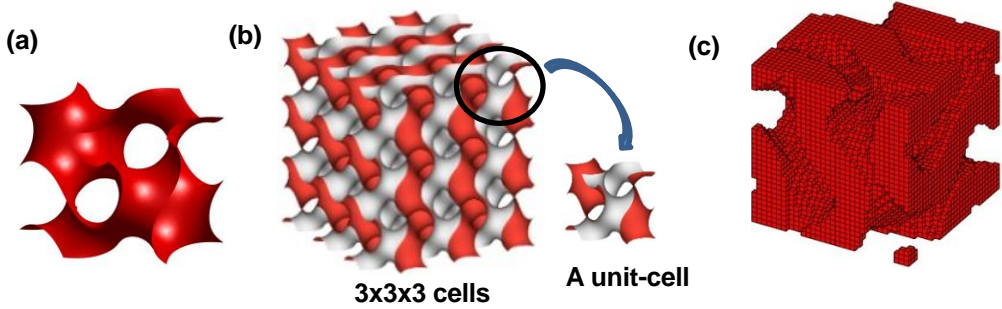

Figure 1. a) Isosurface of the Gyroid with $c = 0,$ b) RUC of a Gyroid along with the periodic structure [4], c) voxelized Gyroid RUC

The isosurface generated from Equations (1) with $c = 0$ is passed to a voxel generation algorithm [44] by passing the vertices' and edges' information of the isosurface. The voxelized RUC (shown in Figure 1c) is generated with a mesh size of 32 voxels in each direction. Each voxel is given a value of 1 (black) if any part of the isosurface edges (obtained from isosurface information) passes through that voxel, else given the value zero (white) to indicate the void space inside the Gyroid. The thin isosurface is thus thickened by the voxel algorithm due to the crisscross connections of the edges-vertices defining the isosurface. The major 2D slices shown in Figure 2 help visualize how the curved interior surfaces of the Gyroid are captured by the voxels. The voxelized RUC has a relative density of 58.7%, obtained by calculating the number of black voxels divided by the total number of voxels (32 x 32 x 32). Using this Gyroid microstructure as the initial design, we can design the optimal Gyroid RUC using a 3D homogenization-based TO approach to maximize either bulk or shear modulus [15]. The broad advantage of using such a voxelized discretization of a smooth surface will eventually be clearer when the concept of CNN is introduced for learning RUC with such a geometry.

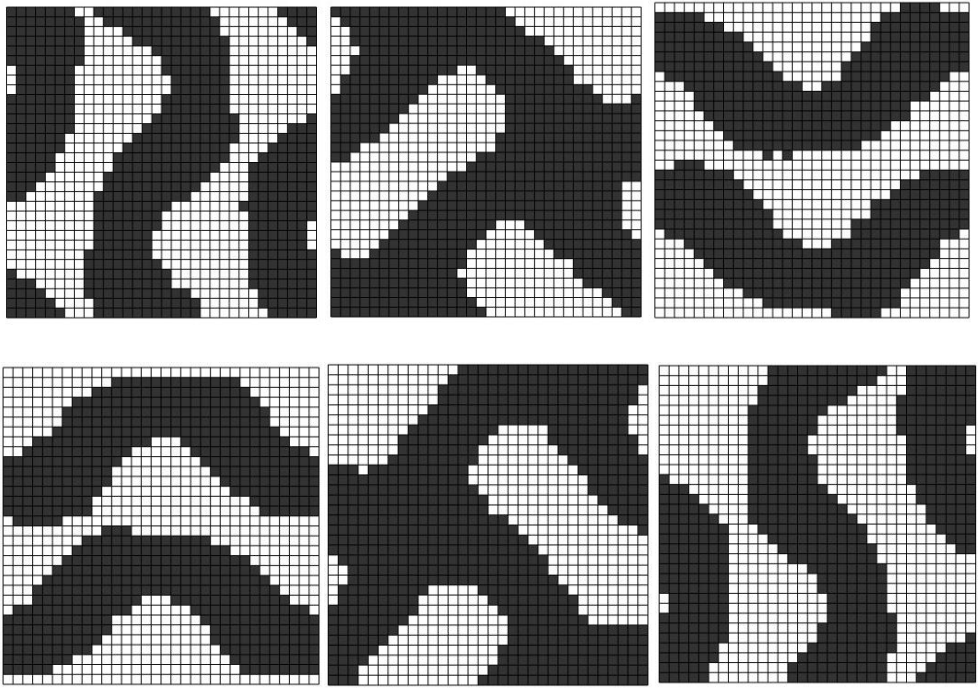

Figure 2: The 1$^{st}$, 4$^{th}$, 8$^{th}$, 16$^{th}$, 24$^{th}$ and 32$^{nd}$ 2D slices of the voxelized Gyroid RUC

**2.2 3D Homogenization based TO**

In this study, we employed a homogenization-based TO approach of microstructure design to design TPMS metamaterials optimized for either maximum bulk or shear modulus [15]. The homogenization method in periodic cellular materials or composites calculates their effective properties [45] using their RUC applying periodic boundary conditions. This effective property of the RUC is then used in the TO algorithm, which maximizes or minimizes a desired objective function. The TO algorithm used for homogenized RUC is the density-based solid isotropic material penalization (SIMP) approach [46], [47] as the proposed voxelized geometry of the unit cell facilitates the calculation of densities with each voxel acting as a finite element in SIMP approach. The energy-based homogenization method is briefly discussed here.

Given a volume of a unit cell, $|Y|$, the homogenized stiffness tensor $E_{ijkl}^H$ is given by volume integrand

$$E_{ijkl}^H = \frac{1}{|Y|} \int_Y E_{pqrs} \varepsilon_{rs}^{A(kl)} \varepsilon_{pq}^{A(ij)} dY \qquad (2)$$

$\varepsilon_{pq}^{A(ij)} = \varepsilon_{pq}^{o(ij)} - \varepsilon_{pq}^{*(ij)}$, $E_{pqrs}$ represents the local stiffness tensor, $\varepsilon_{pq}^{o(ij)}$ is the initial macroscopic strain fields, and $\varepsilon_{pq}^{*(ij)}$ denotes locally varying strain fields. In the case of 3D, there are six prescribed unit test strains $\varepsilon_{pq}^{o(ij)}$ corresponding to independent test strains: $(1,0,0,0,0,0)^T$, $(0,1,0,0,0,0)^T$, $(0,0,1,0,0,0)^T$, $(0,0,0,1,0,0)^T$, $(0,0,0,0,1,0)^T$ and $(0,0,0,0,0,1)^T$. When these unit test strains act on the unit cell, the equilibrium equation with periodic boundary conditions are solved for the unit cell to obtain the unknown strain fields $\varepsilon_{pq}^{*(ij)}$ [15]. The RUC is divided into $N$ finite elements with 6 x 6 element stiffness matrices $\boldsymbol{k}_e$ and $\boldsymbol{u}_e^{A(ij)}$ being element displacements corresponding to $\varepsilon^{o(ij)}$. Hence, the finite element summation of integrand in Equation (2) is written in terms of $\boldsymbol{k}_e$ and $\boldsymbol{u}_e$, in turn, expressed in terms of element mutual energies $Q_{ijkl}^e$ [10] as follows:

$$E_{ijkl}^H = \frac{1}{|Y|} \sum_{e=1}^N \left(\boldsymbol{u}_e^{A(ij)}\right)^T \boldsymbol{k}_e \boldsymbol{u}_e^{A(kl)} = \frac{1}{|Y|} \sum_{e=1}^N Q_{ijkl}^e \qquad (3)$$

The expanded form of this homogenized stiffness tensor [15] is

$$\begin{bmatrix} E_{1111}^H & E_{1122}^H & E_{1133}^H & E_{1112}^H & E_{1123}^H & E_{1131}^H \\ E_{2211}^H & E_{2222}^H & E_{2233}^H & E_{2212}^H & E_{2223}^H & E_{2231}^H \\ E_{3311}^H & E_{3322}^H & E_{3333}^H & E_{3312}^H & E_{3323}^H & E_{3331}^H \\ E_{1211}^H & E_{1222}^H & E_{1233}^H & E_{1212}^H & E_{1223}^H & E_{1231}^H \\ E_{2311}^H & E_{2322}^H & E_{2333}^H & E_{2312}^H & E_{2323}^H & E_{2331}^H \\ E_{3111}^H & E_{3122}^H & E_{3133}^H & E_{3112}^H & E_{3123}^H & E_{3131}^H \end{bmatrix} \qquad (4)$$

The SIMP algorithm in TO is then performed on homogenized RUC. The element densities $\rho_e \in [0,1]$ of each finite element is the design variable and the element Young's modulus constituting $\boldsymbol{k}_e$ in terms of densities is

$$E_e(\rho_e) = E_{min} + (E_o - E_{min})\rho_e^p \qquad (5)$$

where $E_o = 1 \text{GPa}$, solid element Young's modulus and $E_{min} = 1e-9$ GPa, void Young's modulus, introduced to prevent singularity in the stiffness matrix. The penalization factor $p$ is taken here as 5.0. To avoid numerical instabilities of mesh dependence and checker boarding [48], a density filtering approach is adopted which uses filtered densities $\boldsymbol{\rho}$ calculated from pseudo densities $\boldsymbol{\eta}$ for the optimization. The relations between the densities are given below [49]:

$$\boldsymbol{\rho} = \overline{\boldsymbol{W}}\boldsymbol{\eta}$$

$$w_{ij} = \max\left(0, r_{min} - \|\boldsymbol{X}_i - \boldsymbol{X}_j\|^2\right) \qquad (6)$$

$$\overline{w}_{ij} = \frac{1}{\sum_{k=1}^{N_{r_{min}}} w_k} w_{ij}$$

where $r_{min}$ is the filter radius, $\overline{w}_{ij}$ is normalized weight coefficient forming the normalized matrix $\overline{\boldsymbol{W}}$. $\boldsymbol{X}_i$'s are coordinates of centroid of element $i$. The optimization problem can be now stated as

$$\begin{aligned}
\max_{\boldsymbol{\rho}} &\quad f\left(E_{ijkl}^H(\boldsymbol{\rho})\right) \\
\text{such that:} &\quad \boldsymbol{K}\boldsymbol{U}^{A(ij)} = \boldsymbol{F}^{A(ij)} \\
&\quad V(\boldsymbol{\eta}) - V_f \leq 0 \\
&\quad 0 \leq \rho_e \leq 1
\end{aligned} \qquad (7)$$

$K$-global stiffness matrix, $U^{A(ij)}$-global displacements corresponding to strain case(ij), $F^{A(ij)}$- external force vectors, $V(\eta)$- volume fraction got by dividing element volumes with total volume of the domain and this is not to exceed $V_f$, a prescribed limiting value.

The objective function used in this work is maximizing the bulk modulus and shear modulus which is given by [15]

$$f_b\left(E_{ijkl}^H(\boldsymbol{\rho})\right) = \sum_{i,j=1}^{3} E_{iijj}^H \tag{8}$$

$$f_s\left(E_{ijkl}^H(\boldsymbol{\rho})\right) = \sum_{i,k=1}^{3} E_{ijkl}^H (i \neq j \,\&\, k \neq l) \tag{9}$$

The sensitivities are calculated using the adjoint method expressed as

$$\frac{\partial E_{ijkl}^H}{\partial \rho_e} = \frac{1}{|Y|} p \rho_e^{p-1}(E_o - E_{min})\left(\boldsymbol{u}_e^{A(ij)}\right)^T \boldsymbol{k}_o \boldsymbol{u}_e^{A(kl)} \tag{10}$$

where $\boldsymbol{k}_o$ is element stiffness matrix.

The densities can be made to take a 0 or 1 solution with the use of Heaviside filter [50]. This can be particularly useful while dealing with machine learning methods later in the work as a 0-1 morphology is easier to learn than one with intermediate densities. This filtering introduces a Heaviside step function into the density filter using the following smooth function such that physical density $\rho_e =1$ if $\rho_e > 0$ and zero if $\rho_e = 0$:

$$\boldsymbol{\rho}^H = 1 - e^{-\beta\rho} + \rho e^{-\beta} \tag{11}$$

Here, the parameter $\beta$ controls the smoothness of the approximation. When $\beta =0$, the Equation (11) is similar to Equation (6) and as $\beta$ tends to infinity, the approximation approaches a true Heaviside step function. To avoid local minima and to ensure differentiability in the optimization, a continuation scheme is used to increase $\beta$ gradually from 1 to 512, doubling it every 50 iterations or when change between variables in two consecutive design becomes less than 0.01.

The 2D slices of voxels corresponding to the optimized geometry are shown in Figure 3. The voxelized Gyroid RUC are smoothened using the top3d app software [51] and varying relative densities displayed in Figure 4.

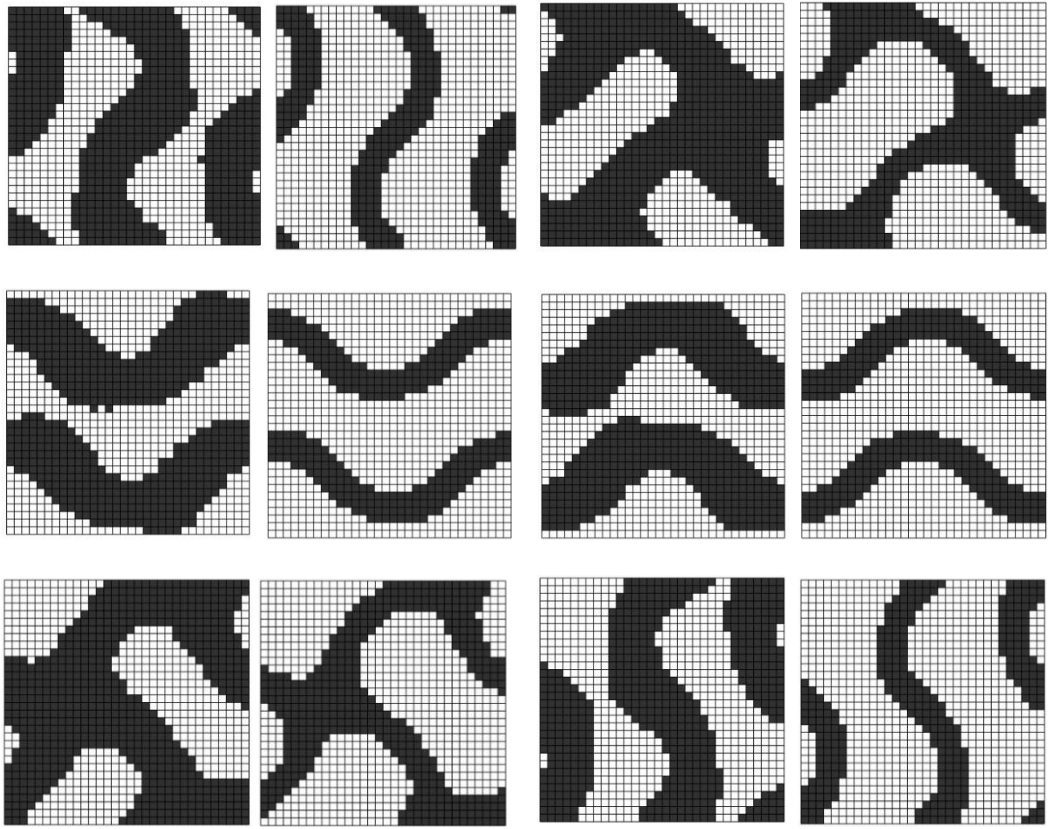

Figure 3: The 1$^{st}$, 4$^{th}$, 8$^{th}$, 16$^{th}$, 24$^{th}$ and 32$^{nd}$ 2D slices of the initial and optimized (34%) voxelized Gyroid RUC

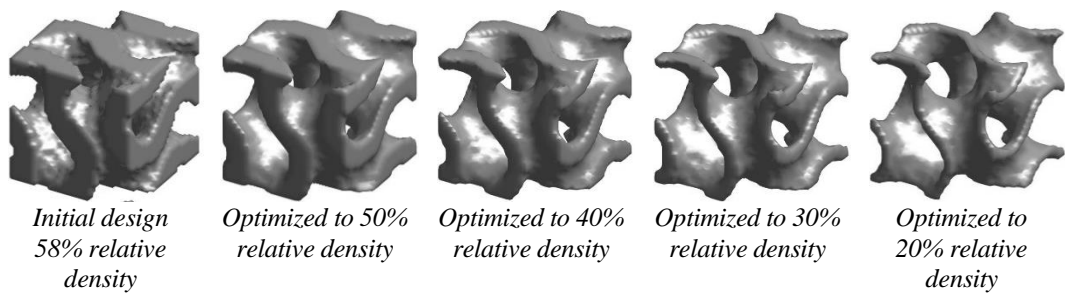

*Initial design 58% relative density*    *Optimized to 50% relative density*    *Optimized to 40% relative density*    *Optimized to 30% relative density*    *Optimized to 20% relative density*

Figure 4: Optimized and smoothened voxelized Gyroid RUC for various relative densities

## 3. SURROGATE DEEP LEARNING MODEL

The homogenization-based TO of the voxelized Gyroid RUC described in the previous section has one lacuna: its computational time. This is indeed one of the major challenges in any 3D unit-cell design; the computational time exponentially grows with the mesh size (number of voxels in this study) of the unit cell. When the objective becomes time-consuming, the 'curse of dimensionality' sets in [52], and it becomes essential to seek alternative ways of determining the objective functions. The method of surrogate modeling [53] appeals in such situations when cheaper alternatives can be employed to perform the objective function evaluations. These models can learn from the information provided from a few optimization runs to replicate the process and consequently be used as a computationally cheap alternative for optimizing unit cell design. Recently, data-driven models have proved effective surrogate models to ease such computationally intensive design through the process of the training-learning algorithm. Among the vast literature on such data-driven models, in this study, we chose the CNN-based model as this class of deep neural networks has proved very successful in image recognition, where images are in the form of pixels in 2D and voxels in 3D. Hence, the broad purpose of Gyroid RUC generation through the voxel algorithm now becomes more meaningful.

This section details the building of a deep learning based model as a surrogate for the topology optimization of the Gyroid microstructure given any volume fraction and filtering radius. The model will predict the optimal 3D Gyroid unit cell, which possesses the maximum bulk/shear modulus for the specified volume fraction and filtering radius, without the need for any traditional topology optimization. This is achieved by training the deep learning CNN model with a few optimized topologies corresponding to different random volume fractions and filtering radii. However, this training requires data to be generated through many topology optimization runs, which is the cost paid for alleviating topology optimization runs later for design.

### 3.1 Design of Experiment (DoE) for Data generation

The flowchart of the workflow is shown in Figure 5. The data required is computationally generated from MATLAB runs of the code containing 3D topology optimization of homogenized properties as described in previous section. As the flowchart indicates, first the isosurface of the Gyroid is generated from Equation (1). Here, we used the value of $c$=0. The isosurface is then voxelized by discretizing the unit cell into 3D finite elements (each element called a voxel), and assigning a density of 1 to each voxel if the isosurface is passing through the voxel and 0 if voxel does not have any part of the isosurface. These voxel densities are used as an initial design for the topology optimization problem, where the bulk or shear modulus is maximized. Two optimization parameters are studied here – volume fraction ($V_f$) in the range of 25%-45% and a filtering radius ($r_{min}$) of the optimization in range of 1.2- 2.5 cm. These parameters are chosen based on previous 2D metamaerial topology optimization studies [40]. To generate data, these two factors are designed in a factorial design and datapoints generated for each pair of values as shown in table 1. The table can be read as the number of the datapoint in center and corresponding value of volume fraction on left side and filtering radius on top of any selected datapoint number. For example, data point 1 has a volume fraction 25% and filter radius of 1.2 cm and so on. Two such tables are created for both the bulk modulus and shear modulus maximization objective.

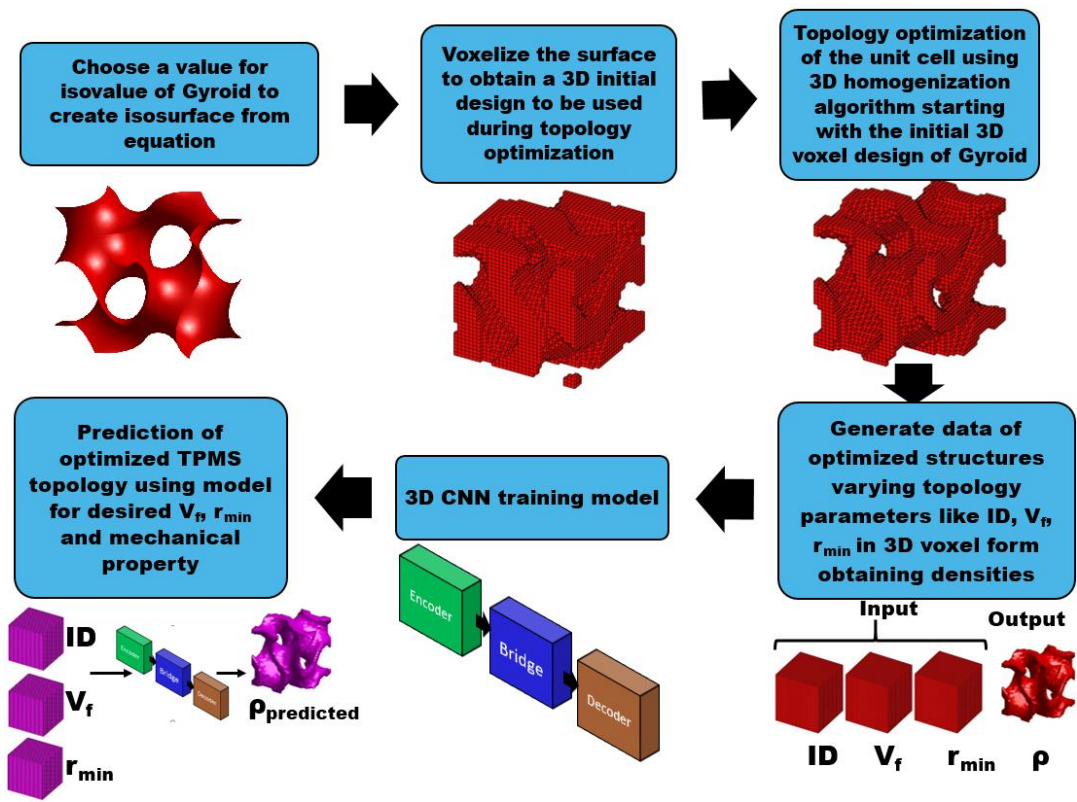

Figure 5. Flowchart of the data generation and prediction process

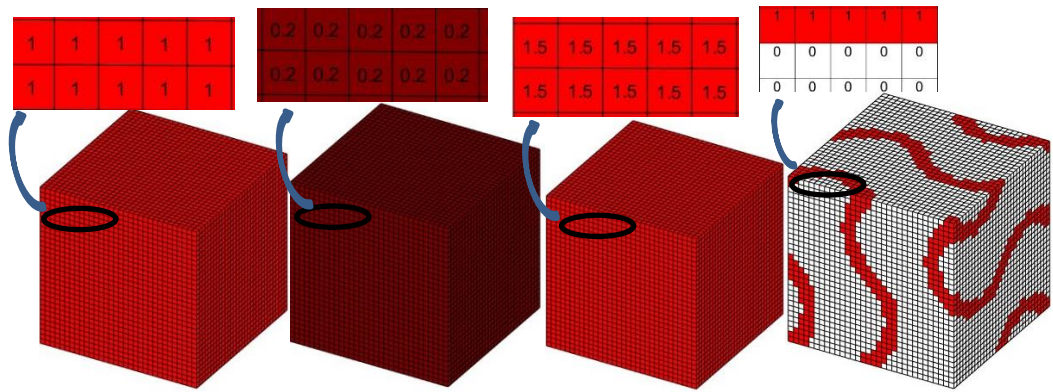

Figure 6. One sample data point -voxels for bulk modulus maximization ID, volume fraction of 20% and filter radius of 1.5 as inputs and corresponding optimized densities as output

Table 1. Datapoints table indicating the values of volume fraction, $V_f$ and filter radius $r_{min}$ for each datapoint

| | | Filter radius(cm) | | | | | |
|---|---|---|---|---|---|---|---|
| | | 1.20 | 1.21 | 1.22 | 1.23 | … | 2.50 |
| Volume Fraction (%) | 25 | 1 | 2 | 3 | 4 | … | 131 |
| | 26 | 132 | 133 | 134 | 135 | … | 262 |
| | . | . | . | . | . | …. | . |
| | . | . | . | . | . | … | . |
| | 45 | 2621 | 2622 | 2623 | 2624 | … | 2751 |

The finite element mesh of 32x32x32 is chosen for the unit cell dimension of 1cm x1 cm x1cm. The choice of mesh will be discussed in detail in Section 4. One sample datapoint is shown in Figure 6 for illustration of how input-output voxel looks like for one set of values of parameters. The choice of the range of optimization parameters is made on the basis that the volume fraction of interest in cellular solids is in range of 25% to 45% while the filtering radius is chosen with a study of different values from 1 to 10. Figure 7 shows some of the shapes of topology optimized result for 40% volume fraction with varying $r_{min}$. We found that large radius filter values lead to reduced effective properties; hence, the maximum filter radius we consider in this study is 2.5 cm. Also the value of $r_{min}$ =1 gives a checkerboard pattern [48] and hence the limits were fixed at 1.2cm to 2.5cm.

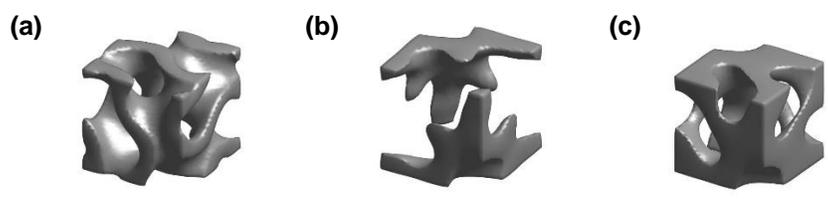

Figure 7. Optimized topologies for volume fraction of 40% for different filtering radii of (a) 3cm, (b) 5cm and (c) 8cm

**3.2 Network architecture**

We extend the CNN model employed to predict optimized 2D metamaterials in a previous work [40] to 3D metamaterials. An encoder-decoder network proposed by ResUnet [54] is used for the model, which is a semantic segmentation neural network taking advantage of both residual learning and U-net [55]. This makes the network include both their strengths. This gives us the motivation to use our pixel-based geometry for learning the property and the related Gyroid RUC geometry (densities) such that for any desired property the model predicts the geometry. The architecture (shown in Figure 8(a)) is similar to a U-Net (called so due to the U-shape of the blocks) with residual blocks instead of neural units as its building block and hence referred to as ResUNet. The architecture can be divided into the encoder part, which encodes the input images into a low-dimensional representation by a series of convolution layers, the decoder which receives the encoded images from the third bridge part, connecting the encoder to the decoder, and constructs back the RUC. The concatenation feature, shown by dotted lines in Figure 8, improves the segmentation accuracy. ResUNet uses batch normalizations (BN), rectified linear units (ReLU), and convolutional layers (Conv), whereas U-Net uses only ReLU and Conv in the building block. Four blocks of Encoder and Decoders are used, and each building block is shown in Figure 8(b). The advantage of this ResUnet over U-net is the concatenation links between the encoder and decoder which helps in preserving features [27].

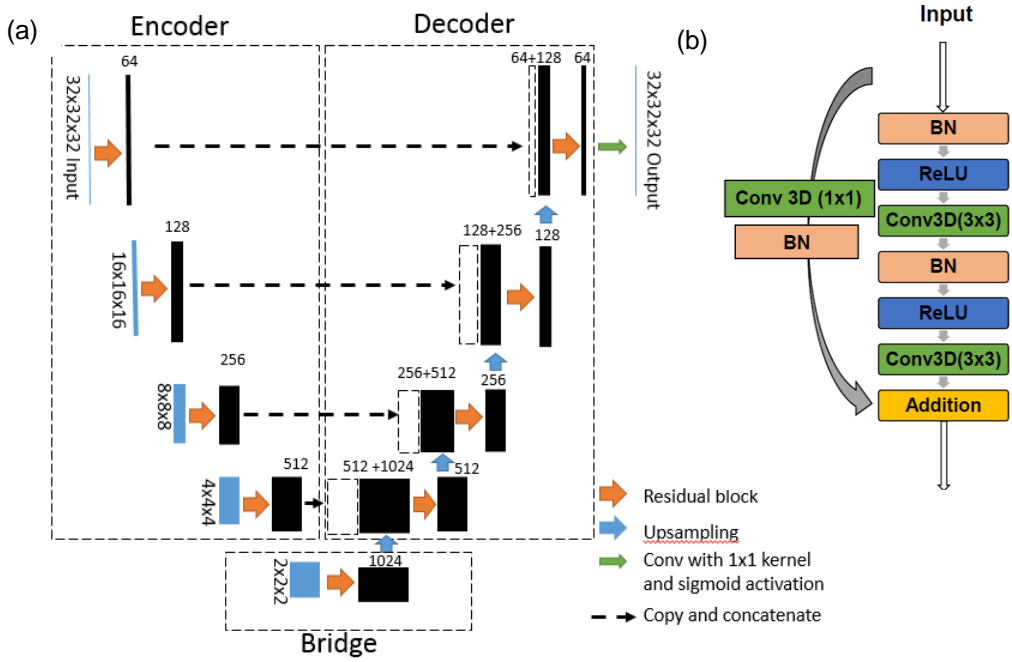

Figure 8. a) ResUNet architecture. The filter sizes are written over each filter(black) while changes in 3D input(blue) is written on sides. b) residual block

The values of topology optimization parameters of $V_f$ and $r_{min}$ and an identifier (ID) for the desired mechanical property (here 1 for maximum bulk modulus and 2 for maximum shear modulus) are converted into 3D matrices (images) assigning same value to all voxels illustrated in Figure 6. The CNN model takes these input 3D images of $V_f$, $r_{min}$ and ID, along with the output 3D image of corresponding topology optimized densities and uses this information to train its weights. Once the training phase is complete, the CNN model is now ready to predict the desired microstructure topology corresponding to any property given to it, as shown in Figure 9.

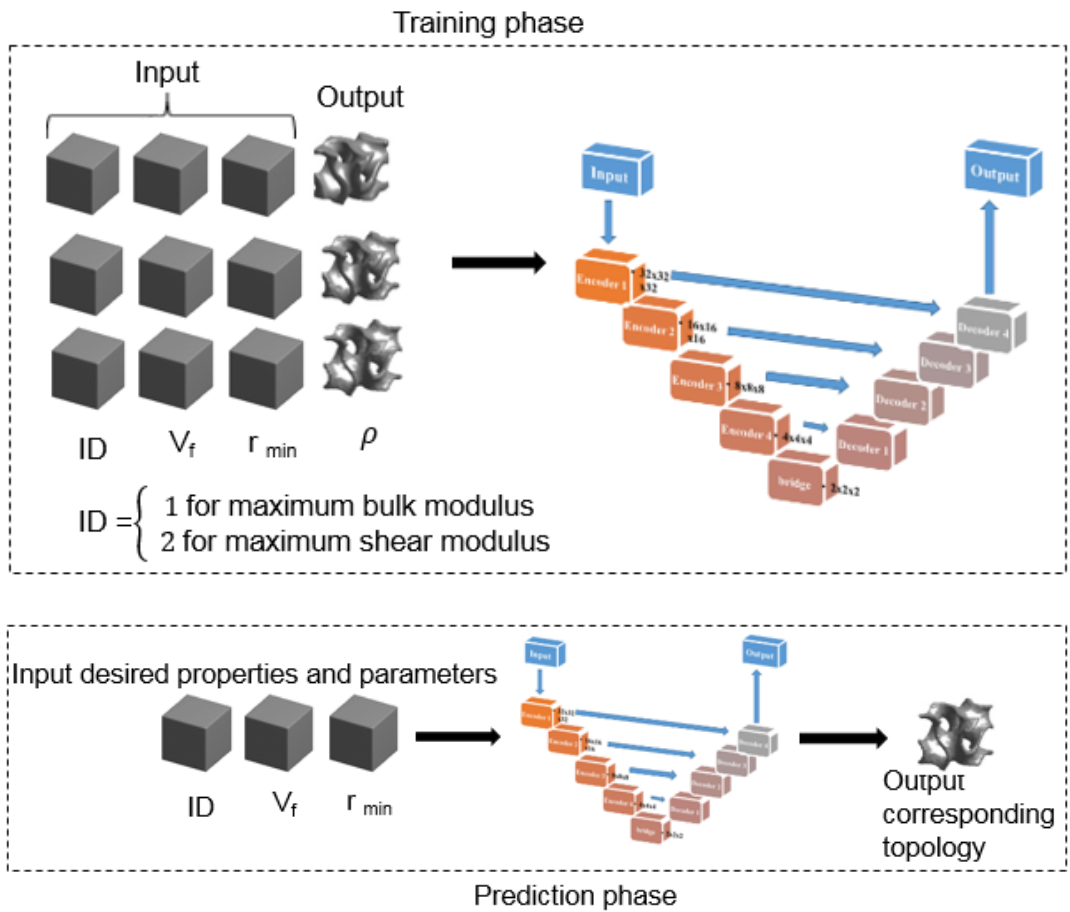

Figure 9. Training and Prediction phases of the proposed 3D CNN network

The dataset generated consists of 2751 datapoints, each datapoint containing an identifier for objective, the volume fraction, filter radius and optimized topology (see Figure 5). The computations were time-consuming and were performed on the IBM HPC with hardware specifications: two 12 core Intel Xeon E5-2695 v2 (Ivy Bridge) CPU, two NVIDIA K20M GPUs, and 264 GB main memory and also on iForge HPC cluster hosted at the National Center for Supercomputing Applications (NCSA) consisting of Intel/Skylake nodes, each with 40 cores and 192 GB of RAM, and a couple of nodes equipped with NVIDIA v100 GPU cards. The CNN model is developed using Keras with Tensorflow backend [19]. The hyper parameters used are: batch-size - 128, learning rate - 0.001, Adam optimizer and 150 epochs. Usually, a large dataset is required for a fast convergence but since the computational cost of topology optimization was high, we started with a small data size to analyze the result. The computational time taken for the entire process is shown in Table 2. For each data point generation on a core on a node of HPC, 2.4 hours are required which multiplied with 2751 data points would have been a herculean task. However, by modifying the MATLAB code using job arrays to split and generate all data points in parallel, as each TO run is independent of each other, data generation is split to 10 data points per MATLAB simulation requiring only 275 runs for the entire data generation. This is achieved with 200 runs on 5 nodes of 40 cores on iForge and 75 runs on 1 node of 26 cores on IBM, in total, taking only 24 hours for the complete data generation. The time for dataset generation on a personal computer and HPC is also compared in Figure 6. As indicated in the table, the deep learning training to calculate the weights and biases takes only 5.5 GPU hours. Once we properly train and validate the deep learning model, the prediction of topologies for new input parameters can be obtained accurately and almost instantly even on a laptop and without any modeling software. This is the greatest advantage of using surrogate deep learning models.

Table 2. Computational time taken on HPC

| Activity | CPU hours | GPU hours |
| --- | --- | --- |
| Data Generation (with 5 nodes of 40 cores on iForge and 1 node of 26 cores on IBM) | 24 | - |
| Deep Learning Training | - | 5.5 |
| Deep Learning Prediction | 0.001 | - |

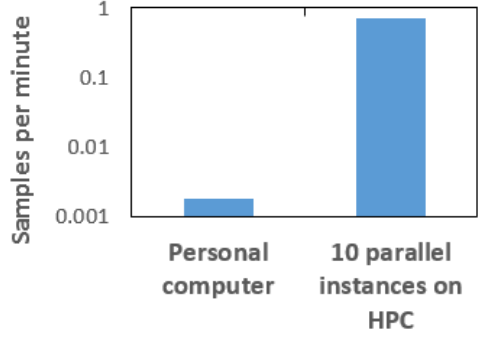

Figure 10. Data generation rates on Workstation versus HPC

**3.3 Model evaluation**

The CNN model was evaluated for its prediction against the ground truth using a mean square error (MSE) metric as the loss function of the model and the mean dice similarity coefficient (DSC) [28] for flattened 3D voxel [19]. The MSE measures how much the predicted topology deviates from the ground truth and smaller values of MSE are preferred. The DSC compares the predicted topology image with the ground truth topology image and gives the measure of how many voxels match in both. So, a higher value for DSC is preferred as that would suggest a higher match between ground truth and prediction.

They are evaluated using the following expressions for $M$ data points, $T$ the ground truth segmentation of input channel $I$ and $O$ the CNN model segmentation,

$$MSE = \frac{1}{M} \sum_{i=1}^{M} \|T(I_i) - O(I_i)\|^2 \qquad (12)$$

$$DSC = \frac{1}{M} \sum_{i=1}^{M} \frac{2|O(I_i) \cap T(I_i)|}{|O(I_i)| + |T(I_i)|} \qquad (13)$$

## 4. ERROR ANALYSIS

Various approximations used during the modeling of the unit cell and its simulations introduce various errors into the model developed for the Gyroid RUC, and this will also affect the CNN modeling of the unit cell. Hence a detailed error analysis is carried out to study all these errors and suggest methods to minimize their effect on the surrogate modeling.

## 4.1 Error in geometry modeling of isosurface

The isosurface is created with different mesh points and number of mesh points can introduce the first discretization error. Figure 11 shows an isosurface with various mesh points, out of which surface is visually best captured by a minimum of 15 points. Mesh sizes of 5, 10, 15, 20, and 32 were used to generate surfaces from which voxelized cubes using 32 voxels (selection criteria discussed in the next section) in each of three directions were generated, and the relative densities of each of these 32 x32 x 32 cube were compared. The difference in relative densities converged after 15 mesh points, indicating the actual relative density of the thickened isosurface is captured. Thus, 15 was chosen as the mesh size to generate the isosurface of c=0, which will be voxelized for analysis.

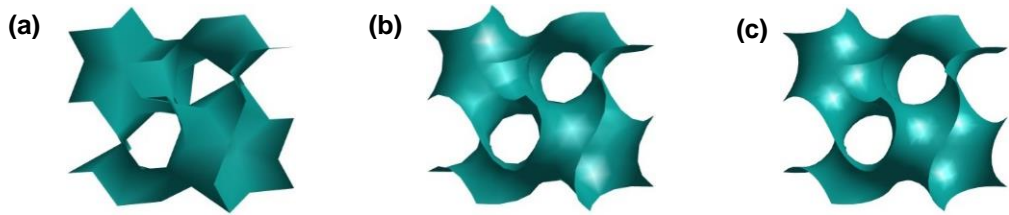

Figure 11. Isosurfaces generated with a)5, b) 10, and c) 15 mesh points.

## 4.2 Error in voxelizing surface geometry

The CNN modeling requires 3D input images with voxels chosen as powers of 2 - either 8, 16, 32, 64, and so on. Hence the voxel size of RUC (or finite element size) was chosen by performing the homogenization-based TO and choosing a mesh size beyond which there was convergence to the homogenized matrix and the compliance of the structure.

Element size of 8 was discarded, as it was less than the mesh points of isosurface (15 chosen in the previous section). Among 16 and 32, since we are analyzing to obtain the best mechanical property, the value of bulk modulus and shear modulus can be studied with both the voxel sizes. The 32 finite element size gave lower compliance and higher bulk modulus and shear modulus value with 60% relative error from those with 16 finite element size, even though it was computationally expensive. Further examination with 64 mesh size showed no improvement in objective functions as those from 32, which was hence the final choice for the voxel size as it considerably improved the mechanical properties from smaller finite element sizes and was also twice the mesh points (15) used to generate the isosurface.

**4.3 Sampling Error**

A small sample size of 2751 data points is bound to create modeling errors in the CNN model, which requires a large dataset for training. This problem is foreseen while choosing data-driven modeling; hence, remedial measures of bootstrapping and cross-validation can resolve such issues. The mean square error metric indicates such lacunae and can be remedied if required. A full factorial DoE is suitable than a random set of data for such a case of low sample size, since we can make sure all the range of values of input parameters are represented in the dataset.

**4.4 Errors in fitting the CNN model**

To prevent the issue of overfitting or underfitting, the loss function (here the mean square error metric) and the dice coefficient metric are studied for both the training and validation set. For this the data is split into training set, testing set and a validation set. Low training set error shows there is no underfitting but a low validation set error is also required to show that overfitting has not occurred. Hence both these errors are monitored. In addition, low testing set error will show a low generalization error.

5. **RESULTS AND DISCUSSION**

The optimized topology dataset for objective functions of bulk modulus and shear modulus is shown in Figure 12. As the filtering radius increases, the objective function decreases, and with a very low filtering radius, the optimization did not converge for low volume fractions. Few topologies from the data sample corresponding to the maximum bulk objective are shown for different values of volume fraction and filter radius (2-6 in Figure 12). Topologies corresponding to a very low filter radius show hollow sections in the Gyroid (1in Figure 12). For example, for a 25% volume fraction, topology corresponding to a filter radius of 1.4cms for bulk modulus objective and 1.2 for shear modulus objective has hollow parts in their topology, which is smoothened out when the filter radius is 1.5cms.

For maximizing objective functions, a filtering radius value of around 1.5cms was ideal for low volume fractions and 1.3 for higher volume fractions. The higher volume fraction led to higher objective values, as expected. Few combinations of volume fraction and filtering radii did not converge even. The gaps in the surface show values for which the topology optimization did not converge after 1000 iterations and hence were discarded from the dataset. They are indicated by gaps in the surface in the figure. Hence, out of 2751 datapoints, the final training dataset consists of 2597 datapoints maximized for bulk modulus and 2741 datapoints maximized for shear modulus. Each datapoint would include the identifier indicating the objective (maximizing shear or bulk), volume fraction value, the filtering radius value, and the 32x32x32 values of densities. Both the objective function datapoints were mixed and shuffled for the training dataset. The data was split into 90% training data, 5% validation data, and 5% testing data.

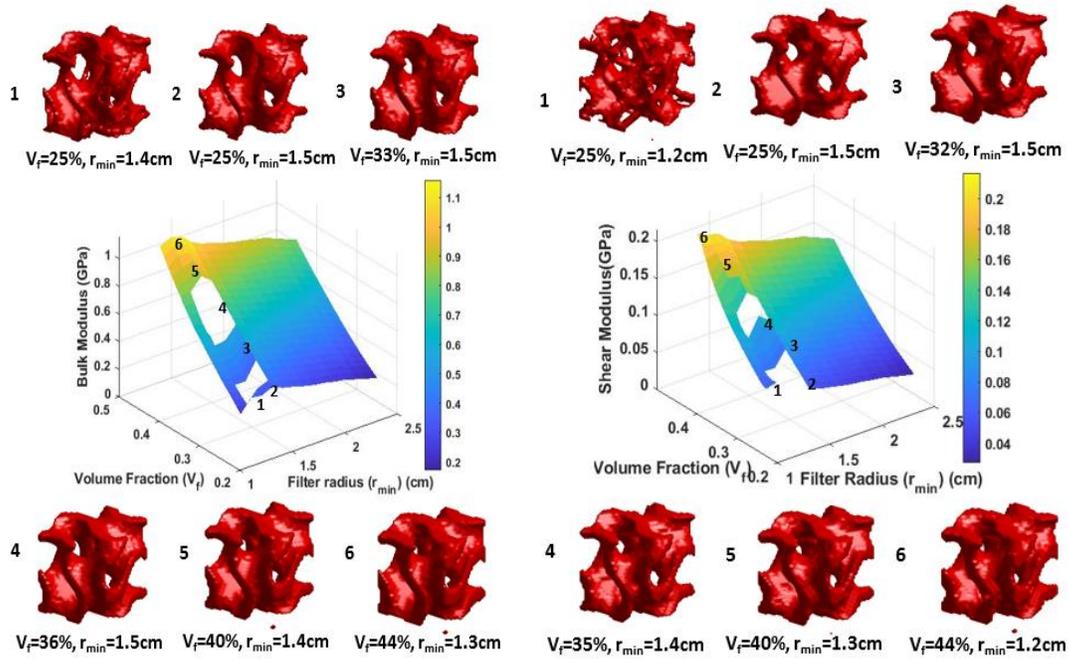

Figure 12. Topology optimized objective function surfaces for different combinations of volume fractions and filter radii.

The deep learning method used for learning the 3D Gyroid topologies optimized for maximum bulk modulus and maximum shear modulus is tested for its effectiveness. The measure of effectiveness is indicated by the loss function adopted for the model shown in Figure 13(a). The convergence of MSE occurs around 100 epochs, even with a small dataset. The mean DSC history (Figure 13(b)) also indicates a 95% match between predicted and ground truth topologies around 100 epochs. This was also possible partly because of the use of the Heaviside filter [50] in topology optimization which pushes the density values to either 0 or 1 and this has helped the CNN model learn the density image faster as either black or white rather than having intermediate densities. This improved the mean DSC of the dataset. Figure 13 also indicates that the difference between the validation and training error is very small indicating that there is no overfitting or underfitting. Few comparisons of ground truth and CNN predicted topologies obtained from the testing set, corresponding to filter radius of 1.2, 1.5, 1.8, 2.2, 2.5 for volume fractions 25%, 35% and 45% are shown in Figure 15 and Figure 16. As is visually noticed, the low filter radius for low volume fractions gave discontinuities in topologies which were not efficiently learnt by the CNN model while it performed exceptionally well for higher filtering radii for all volume fractions. For the testing set, the MSE was found to be 0.0079 showing a low generalization error. The mean deviation of volumes of predicted structures from the ground truth was evaluated for this test set and found to be 0.24%. Among this, the highest deviation of volume was shown by a structure with 29% volume fraction and 1.35 cm filter radius optimized for shear modulus with an absolute error in volume as 1.73%. The lowest deviation in volume showed by a 40% structure volume fraction and 1.41 cm filter radius optimized for bulk modulus having error as 7.6e-4%.

For a better understanding of the matching images, the $1^{st}$, $8^{th}$, $16^{th}$, $23^{rd}$ and $32^{nd}$ 2D slices among the 32 slices of the 3D image are separately visualized for both the objectives for a 35% volume fraction and filter radius of 2 cm is shown in Figure 16.

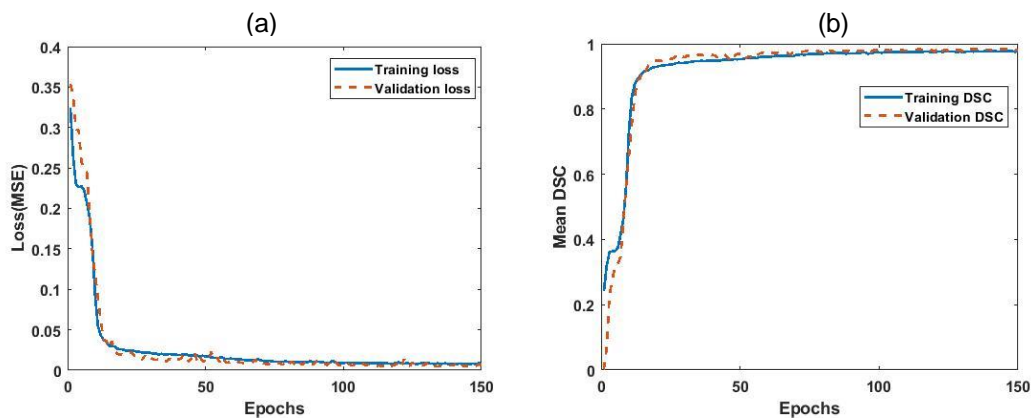

Figure 13. a) MSE convergence and b) mean dice similarity coefficient, against number of epochs.

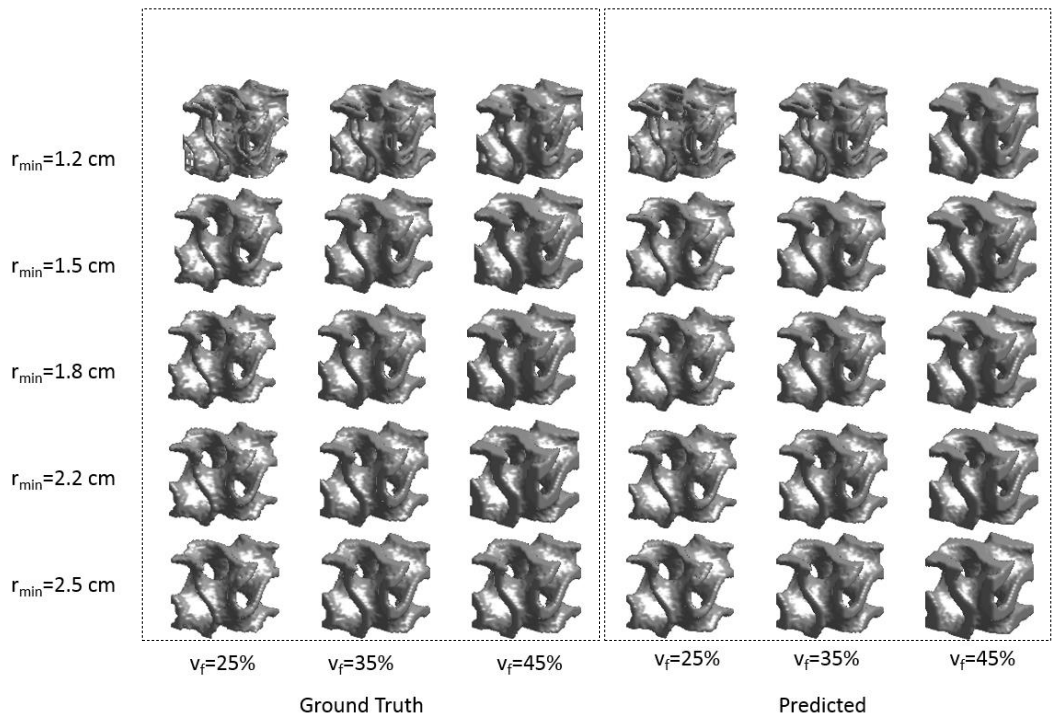

Figure 14. Ground truth (left) and predicted (right) topologies for different combinations of volume fractions and filtering radii with maximum bulk modulus

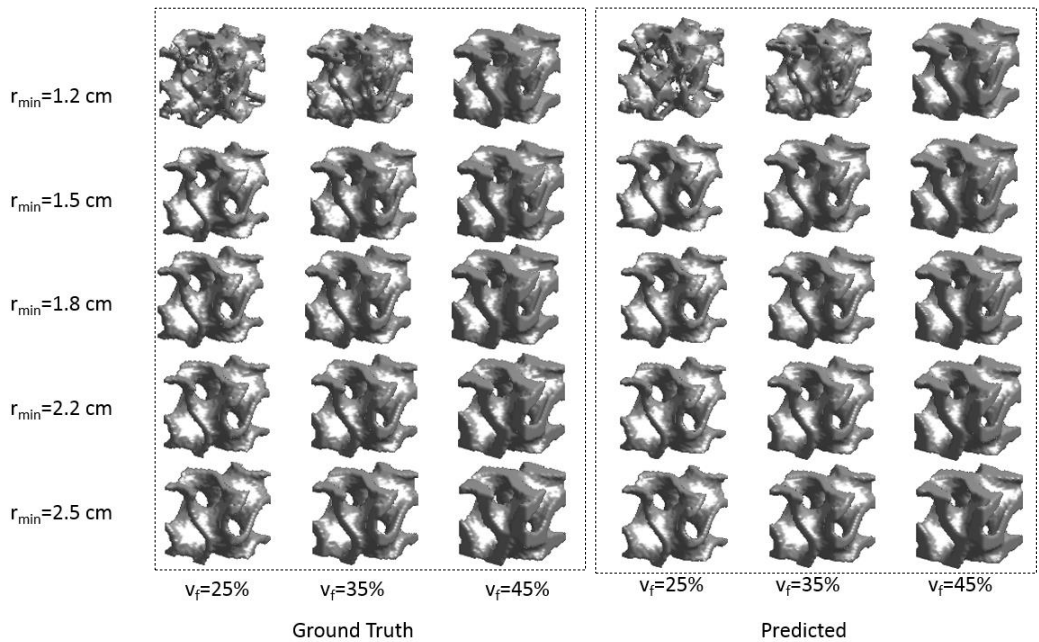

Figure 15. Ground truth (left) and predicted (right) topologies for different combinations of volume fractions and filtering radii with maximum shear modulus

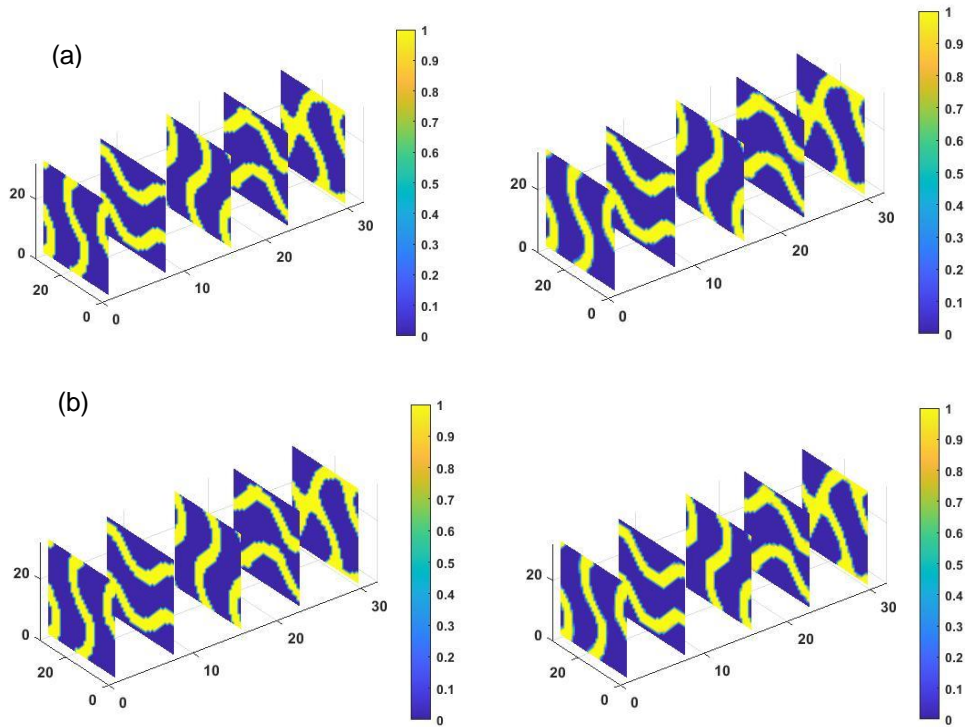

Figure 16. 1st, 8th, 16th, 23rd and 30th 2D slice contours of densities in ground truth (left) and predicted (right) topologies for 35% volume fraction and filter radius of 2cm for a) maximum bulk modulus and b) maximum shear modulus

## 6. APPLICATIONS

The future applications of the conducted research can further emphasize the significance of the work studied. The 3D TO of a large mesh size is time-consuming, and generating a dataset of many such TO runs is expensive. A bigger purpose and application should be the aim while performing such an exercise. TPMS-based porous structures design for lightweight mechanical structures, heat exchangers, and biomaterials is researched by studying the RUC design and then using the optimal unit cell to generate periodic macrostructures. Nevertheless, the structural and material optimization of the macrostructure may not be possible by using an optimal microstructure alone. Towards this, integrated topology optimization called concurrent TO, which optimizes the microstructure material distribution at the same time when macrostructure properties are optimized, is studied widely [56],[57] . This multiscale TO method is found to be even more computationally intensive, in the order of the finite element mesh size of microscale multiplied by the finite element size of macroscale. When we develop a model for determining the optimal microstructure corresponding to any desired mechanical property and TO parameters instantaneously, this model can be plugged into the macro-analysis of such structures to prevent any such concurrent topology optimization of structure at micro-macro scales. The concurrent TO involves the following processes [56]:

Find $\rho_M^i, \rho_m^i (i=1,2,\ldots,N_M; j=1,2,\ldots,N_m)$

Min: $C(\rho_M, \rho_m)$

$$\text{such that:} \quad \boldsymbol{K(D_M)U_M^{A(ij)} = F_M^{A(ij)}, K(D_m)U_M^{A(ij)} = F_m^{A(ij)}} \tag{14}$$

$$V_M(\rho_M) - V_{fM} \leq 0, V_m(\rho_m) - V_{fm} \leq 0$$

$$0 \leq \rho_M^i \leq 1, 0 \leq \rho_m^i \leq 1$$

where $C$ is the structural compliance, $M$ index refers to macrostructure and $m$ to microstructure. $\rho_M, \rho_m$ are relative densities and $\boldsymbol{D_m}$ and $\boldsymbol{D_M}$ are stiffness tensors of microstructure and macrostructure calculated similar to Eq. (5) as

$$D_M = [E_{min} + (E_o - E_{min})\rho_M^p]D^H$$
$$D_m = [E_{min} + (E_o - E_{min})\rho_m^p]D^0 \tag{15}$$

where $\boldsymbol{D^0}$ is constitutive matrix of the material and $\boldsymbol{D^H}$ is homogenized stiffness tensor of the microstructure optimized by TO. It is this TO which is modeled in our study with a CNN model and hence can be used as an alternate in the concurrent TO to reduce the overall computational cost of TO of entire macrostructure. In this study we calculate the bulk modulus and shear modulus from $\boldsymbol{D^H}$ which is one of the input parameters. Instead, for the concurrent TO, the $\boldsymbol{D^H}$ can be the desired property to be attained and the model built in the same procedure as detailed in this study. Hence a major savings in computational time is achieved on macrostructure design as the microstructure design is predicted by CNN model instantaneously in each iteration based on TO parameters. The similar applications where the macrostructure is analyzed with computationally intensive FE or CFD models for static [58] or dynamic analysis and which involves the property optimization of microstructure can also use the advantage of this model. The authors are extending this approach to such applications which is the future scope of this research.

7. **CONCLUSION**

The paper introduces a 3D CNN-based model for topology optimization of Gyroid TPMS unit cells. Three novel ideas are presented in the paper – 1) A voxelized algorithm for unit cell design of the 3D Gyroid unit cells, 2) homogenization-based 3D TO to achieve maximum bulk modulus or shear modulus for the desired volume fraction and filtering radii of this microstructure and 3) 3D CNN for 3D TO. To alleviate the computational burden caused by time-consuming 3D TO, a 3D surrogate CNN model with an encoder-decoder type architecture, used in segmentation modeling, is used to learn the topology of the RUC. It was observed that the model could almost instantly imitate a similar pattern in the topology of the Gyroid with very few datapoints. Moreover, the model was robust in both the accuracy of prediction and prediction time. Hence this CNN model could be used effectively, even on a laptop, for performing quality TO, which is otherwise unthinkable even on a powerful workstation or cluster. This work shows promises in employing surrogate deep learning based models for a drastically accelerated unit cell design of 3D metamaterials involving computationally extensive TO, including multiscale metamaterial design.

## ACKNOWLEDGEMENT

The authors would like to thank the National Center for Supercomputing Applications (NCSA) Industry Program, the Center for Artificial Intelligence Innovation and also the Research Computing team at Khalifa University. This publication is based upon work supported by the Khalifa University under Awards No. RCII-2019-003.